\documentstyle[12pt,epsfig]{article}
\newcommand{\be}[1]{\begin{equation} \label{(#1)}}
\newcommand{\ee}{\end{equation}}
\newcommand{\ba}[1]{\begin{eqnarray} \label{(#1)}}
\newcommand{\ea}{\end{eqnarray}}
\newcommand{\nn}{\nonumber}
\newcommand{\rf}[1]{(\ref{(#1)})}

\def\pmb#1{\setbox0=\hbox{#1}%
  \kern-.015em\copy0\kern-\wd0
  \kern.03em\copy0\kern-\wd0
  \kern-.015em\raise.0233em\box0 }
\def \znbb {0\nu\beta\beta}

\begin{document}
%\preprint{USM-TH-132}
%
%
\begin{center}
   {\Large\bf Proton Stability in Leptoquark Models}\\[5mm]
Sergey Kovalenko\footnote{e-mail: kovalenko@fis.utfsm.cl},
Ivan Schmidt\footnote{e-mail: ischmidt@fis.utfsm.cl}\\[1mm]
{\it Departamento de F\'\i sica, Universidad
T\'ecnica Federico Santa Mar\'\i a, Casilla 110-V, Valpara\'\i so, Chile}\\
\end{center}
\bigskip
%
%
%%%%%%%%%%%%%%%%%%%%%%%%%%%%%%%%%%%%%%%%%%%%%%%%%%%%%%%%%%%%%%%%%%%%%%%%%%%
%%              
%%%%%%%%%%%%%%%%%%%%%%%%%%%%%%%%%%%%%%%%%%%%%%%%%%%%%%%%%%%%%%%%%%%%%%%%%%%
\begin{abstract}
We show that in generic leptoquark (LQ) extensions of the standard model
lepton and baryon numbers are broken at the level of renormalizable operators. 
In particular, this may cause fast proton decay unless the leptoquarks are heavy enough.
We derive stringent bounds for the 1st generation LQ masses and couplings 
from the proton stability constraints.
\end{abstract}
%%%%%%%%%%%%%%%%%%%%%%%%%%%%%%%%%%%%%%%%%%%%%%%%%%%%%%%%%%%%%%%%%%%%
%\bigskip
%\pacs{12.60.-i, 13.30.-a,  14.80.-j}  
\vskip 3mm

Leptoquark models \cite{ps,Lagr1} remain an attractive possibility for 
new physics beyond the Standard Model (SM) admitting non-SM particles, leptoquarks (LQ), 
with masses at the electroweak scale. LQs are vector or scalar particles carrying both 
lepton and baryon numbers and, therefore, having a distinct experimental signature.
%For these reasons, LQs are subject of  present and near future experimental searches
%\cite{Global-1,Global-2}. 
For these reasons searching for LQs is a promising subject 
for present and near future experiments \cite{Global-1,Global-2}.

Theoretical motivation for LQ models usually refers to the low energy 
limit of some more fundamental theory associated with an energy scale much higher than 
the electroweak scale. In the literature it has been argued that the origin of LQ models may 
reside in grand unified theories (GUT) \cite{GUT}, \cite{sstr}, models of 
extended technicolour \cite{tech}-\cite{pi_decay}, composite models \cite{compos} and some other
high-energy scale model. 
However the arguments within this framework in favor of light LQs with 
masses of the order of the electroweak scale are quite vague. 
Nevertheless the generic structure of LQ models can be completely determined by the
symmetry with respect to the SM gauge group \cite{Lagr1,HKK}. In this way LQ models with
light LQs can be studied without referring to their high energy scale origin.

Adopting this approach we examine the question of Lepton (L) and Baryon (B) 
number conservation in LQ models. We show that in the LQ Lagrangian there 
exist renormalizable $\Delta L = 1$ and $\Delta B =1$ terms which may affect the conventional LQ 
phenomenology. In particular, the previously overlooked in the literature $\Delta B = 1$ 
terms may cause fast proton decay unless the LQs are heavy enough. Thus the proton 
stability constraint casts lower limits on the LQ masses. We show that these limits are 
more stringent than those existing in the literature \cite{Global-1,Global-2}. 
%and weaker dependent on the LQ coupling constants. 

Here we construct the interaction Lagrangian of a generic leptoquark model
retaining all the renormalizable couplings invariant under the SM gauge group 
$SU(3)_c\otimes SU(2)_L\otimes U(1)_Y$.
We separate the LQ interaction Lagrangian in the following three parts
\ba{Int-Lagrangian}
{\cal L}_{LQ-int} = {\cal L}_{LQ-l-q} + {\cal L}_{LQ-H} + {\cal L}_{LQ},
\ea
corresponding to LQ-lepton-quark, LQ-Higgs interactions and LQ self-interactions.  

The LQ-lepton-quark interaction terms ${\cal L}_{LQ-l-q}$, which fix the LQ field content, 
were constructed in Ref.  \cite{Lagr1} for the scalar $S$ and vector $V_{\mu}$ leptoquarks.
Here we show only the scalar LQ interactions
\ba{S-l-q}\nn
{\cal L}_{S-l-q} &=&
\lambda^{(R)}_{0}\cdot \overline{u^c} P_R e \cdot S_0^{R\dagger} +
%%%%%%%%%%%%%%%%%%%%%%%%%%%%%%%%%%%
\tilde\lambda^{(R)}_{0}\cdot \overline{d^c} P_R e \cdot
\tilde{S}_0^{\dagger} 
%%%%%%%%%%%%%%%%%%%%%%%%%%%%%%%%%%%
+ \lambda^{(R)}_{1/2}\cdot \overline{u} P_L l
\cdot {S}_{1/2}^{R\dagger} +
%%%%%%%%%%%%%%%%%%%%%%%%%%%%%%%%%%%
\tilde\lambda_{1/2}\cdot \overline{d} P_L l \cdot
\tilde{S}_{1/2}^{\dagger} + \\ 
%%%%%%%%%%%%%%%%%%%%%%%%%%%%%%%%%%%
&+& \lambda^{(L)}_{0}\cdot \overline{q^c} P_L i \tau_2 l \cdot
S_0^{L\dagger} +
%%%%%%%%%%%%%%%%%%%%%%%%%%%%%%%%%%%
\lambda^{(L)}_{1/2}\cdot \overline{q} P_R i \tau_2 e \cdot
S_{1/2}^{L\dagger} 
%%%%%%%%%%%%%%%%%%%%%%%%%%%%%%%%%%%
+ \lambda_{1}\cdot \overline{q^c} P_L i \tau_2
\hat{S}_1^{\dagger} l + h.c.,
\ea
where $P_{L,R} = (1\mp\gamma_5)/2$;  $q$ and $l$ are the quark
and lepton doublets; $S_j^{i}$ are the scalar LQs with weak 
isospin j=0, 1/2, 1, coupled to left-handed ($i = L$) or right-handed ($i = R$)
quarks respectively (for a discussion on chiral couplings see Ref. \cite{HKK}).  
The LQ quantum numbers are listed in Table 1.
For LQ triplets $\Phi_1 = S_1, V_1^{\mu}$ we use the notation
$\hat{\Phi}_1 = \vec\tau\cdot\vec{\Phi}_1$.
 
To obey the stringent constraints from FCNC processes it is usually 
assumed that the LQ couplings are generation ``diagonal", i.e. they couple only to
a single generation of leptons and quarks. This implies that there exist
three generations of LQs with the assignments of Table 1. 
In general these couplings could involve all fermion and LQ generations. 

The renormalizable LQ-Higgs interaction terms for the scalar and vector LQs 
have been constructed in Ref. \cite{HKK}. Again, for brevity we show only the scalar 
LQ interactions:
\ba{LQ-Higgs}
{\cal L}_{LQ-H}&=& h^{(i)}_{0} H i\tau_2 \tilde{ S}_{1/2}\cdot S_0^{i}
+  h_{1} H i\tau_2 \hat S_{1}\cdot \tilde{S}_{1/2} 
+ Y_{1/2}^{(i)} \left(H i\tau_2 S_{1/2}^{i}\right)\cdot
\left(\tilde{S}_{1/2}^{\dagger} H\right) +\\ \nn
&+& Y_{1} \left(H i\tau_2 { \hat S}_{1}^{\dagger} H\right)\cdot
\tilde{S}_{0} 
+ \kappa^{(i)} \left(H^{\dagger} \hat S_{1} H\right)\cdot
S_{0}^{i\dagger}
- \left(M^2_{\Phi} -
g_{\Phi}^{(i_1i_2)} H^{\dagger} H\right) \Phi^{i_1\dagger} \Phi^{i_2}.
\ea
Here $H$ is the SM $SU(2)_L$-doublet Higgs field.
$\Phi^{i}$ is a cumulative notation for all the leptoquark fields with
$i = L,R$ (the same for $i_{1,2}$).

The new, previously overlooked part of the LQ interaction Lagrangian, corresponds
to the LQ self-interaction terms.  For brevity we write down this part in the form of 
SM group singlet products of the LQ representations 
\ba{LQ-self}
{\cal L}_{LQ} &=&
\mu_{0}^{i} (S_0^{{i}{\dagger}}\times \tilde{S}_{1/2}\times \tilde{S}_{1/2})+
\tilde{\mu}_{0}^{i} (\tilde{S}_0^{{\dagger}}\times {S}_{1/2}^{i}\times \tilde{S}_{1/2})
+ \mu_{1} (S_1^{{\dagger}}\times \tilde{S}_{1/2}\times \tilde{S}_{1/2}) +\\ \nn
&+& g_{0}^{i} D\cdot (V_0^{{i}{\dagger}}\times \tilde{V}_{1/2}\times \tilde{V}_{1/2})
+ \tilde{g}_{0}^{i} D\cdot  (\tilde{V}_0^{{\dagger}}\times {V}_{1/2}^{i}\times \tilde{V}_{1/2})+
g_{1} D\cdot  (V_1^{{\dagger}}\times \tilde{V}_{1/2}\times \tilde{V}_{1/2}) + \\ \nn
&+& \eta_{\alpha\beta} (\Phi_{\alpha}\times \Phi_{\alpha})\cdot 
(\Phi_{\beta}\times \Phi_{\beta}) 
+ \eta^{ij}_{0} (\Phi_0^i\times \tilde{\Phi}_0^{\dagger}\times \Phi_{1/2}^j\times
\tilde{\Phi}_{1/2}^{\dagger})
+ \eta^{ij}_{1} (\tilde{\Phi}_0^{\dagger}\times \tilde\Phi_{1/2}^{\dagger}\times 
{\Phi}_{1/2}^{i}\times {\Phi}_{1}^j),
\ea
where $\Phi = S,V$ and the superscripts $i,j = L,R$ are LQ chirality indexes defined 
in Eq. \rf{S-l-q}. The subscripts $\alpha, \beta$ in the 1st quartic term denote all
the types of vector and scalar LQs. The parameters  $\mu_a^b$ and $g_a^b, \eta_a^{ij}$  
are dimensionful and dimensionless parameters of the model respectively. 
The SM gauge group covariant derivative  
$D_{\mu}$, defined in the standard way, acts on all the LQ fields $V$ in brackets. 
Here we suppressed the Lorentz indexes of the vector LQs which must be contracted in 
all the possible pairs to form Lorentz scalar products. 
Thus the above indicated vector LQ couplings represent  
groups of terms with different coupling constants for each term. 
For example, the first group of vector LQ couplings contains the term 
$V_{0\ \nu}^{i\dagger}\times \tilde{V}_{1/2}^{\mu}\times (D_{\mu} \tilde{V}_{1/2}^{\nu})$
which has its own coupling constant $\mu_{0}^{i(1)}$ {\it etc.} 

Let us check the lepton and baryon number properties of the terms in the LQ Lagrangian. 
All the LQ-lepton-quark interaction terms ${\cal L}_{LQ-l-q}$ conserve both L and B 
numbers by construction and, therefore, are  $\Delta L = \Delta B = 0$ operators. In the 
sector of the LQ-Higgs interactions ${\cal L}_{LQ-H}$ the quartic terms are also 
$\Delta L = \Delta B =0$ operators while the trilinear terms violate lepton number 
but conserve baryon number so that in this case $\Delta L = 2,\  \Delta B =0$. 
In the LQ self-interaction part ${\cal L}_{LQ}$ again
the quartic terms are $\Delta L = \Delta B =0$ operators while the trilinear terms
violate both lepton and baryon number as $\Delta L = - \Delta B = 1$. 

The trilinear $\Delta L = 2,\  \Delta B =0$ terms in the LQ-Higgs sector contribute 
to lepton number violating processes, such as neutrinoless double beta decay $(\znbb)$ 
\cite{HKK}, as well as to the Majorana neutrino mass matrix. Thus from the experimental 
limits on the rates of these processes and neutrino masses one can deduce the corresponding 
limits on the LQ model parameters.

The trilinear $\Delta L = - \Delta B = 1$ LQ self-interaction terms in 
Eq. \rf{LQ-self} can induce proton decay and, therefore, the existing 
stringent constraints on proton stability may produce valuable lower bounds on the LQ masses. 
These terms, combined with the LQ-lepton-quark couplings from Eq. \rf{S-l-q}, generate 
at low energies $\Delta L = - \Delta B = 1$ lepton-quark contact effective operators.  
For example, the 1st LQ self-interaction term in Eq. \rf{LQ-self} generate the following
dim=9 operator 
\ba{example}
\hat{O}^{(9)} = \lambda_0^i\ \tilde\lambda_{1/2}^2
\cdot \left(\frac{\mu_0^i}{M_{S}^6}\right) (\bar\nu\ P_R d) 
(\bar{e} \ P_R d)\ (\bar{u^c}\ P_R e),
\ea
where $M_{S}$ is the typical mass of scalar LQs.
This and the other possible $\Delta L = - \Delta B = 1$  terms can induce proton decay 
in the channels:
\ba{decay} 
p\ \rightarrow \ e^- e^+ \nu \pi^0 \pi^+,\  e^- \bar\nu \nu \pi^+ \pi^+,\  e^- e^+ \nu \pi^+,\ 
\nu \nu \bar\nu \pi^+.
\ea
The corresponding proton lifetime $\tau_p$ can be estimated in the usual way. In this estimate 
we assume that all the dimensionless coupling constants in Eq. \rf{S-l-q} are of the same order
of magnitude $\lambda_i^j \sim \lambda$ and that all the trilinear scalar LQ self-interaction 
terms in Eq. \rf{LQ-self} have the same mass scale 
$\mu_0^i \sim \tilde{\mu}_0^i \sim \mu_1 \sim \mu$. We also assume that there is no strong 
cancellation between different contributions to the proton lifetime. With this assumptions 
we obtain for the contributions of the scalar LQ  the following estimate
\ba{estim-scal}
\tau_p^{-1} = \kappa \cdot \lambda^6 \cdot \left(\frac{1 \mbox{GeV}}{M_{S}}\right)^{12}
\cdot \left(\frac{\mu}{1 \mbox{GeV}}\right)^{2}\cdot m_p.
\ea
Here $\kappa$ is the dimensionless phase space factor.
The typical energy scale of proton decay is the energy released in this reaction. 
In the above formula the energy scale is set by the proton mass
$m_p$. The difference between $m_p$ and the actual energy released in a specific channel 
of proton decay is absorbed in the factor $\kappa$. From the existing 
lower experimental bound on the proton life time 
(channel independent) $\tau_p^{exp} \geq 1.6 \times 10^{25}$ years \cite{PDD} we obtain
for the scalar LQ mass
\ba{M-LQ}
M_{S} \geq &\kappa^{1/12}& \lambda^{1/2} \left(\frac{\mu}{1 \mbox{GeV}}\right)^{1/6}
\cdot 10\ \mbox{TeV} 
\approx \lambda^{1/2} \left(\frac{\mu}{1 \mbox{GeV}}\right)^{1/6}
\cdot 10\ \mbox{TeV}.
\ea 
Here we put $\kappa^{1/12} \approx 1$ which is a good approximation taking into account the 
very small exponent of the phase space factor. For the same reason this result is weakly dependent
on the simplifying assumptions made before Eq. \rf{estim-scal}.  
In order to reduce the number of free parameters in the above formula and deduce more 
information on the lower bound for the LQ mass one needs some assumptions on 
the possible values of the coupling
constants $\lambda$ and on the mass scale $\mu$ of the trilinear operators in Eq. \rf{LQ-self}.
If the theory is in the perturbative regime one may assume that the 
dimensionless couplings $\lambda$ obey the condition 
$\lambda^2/(4 \pi)\leq 1$. Assuming further that 
the LQ model originates from some GUT scenario one may also think that 
these couplings are of the order of the Higgs-fermion Yukawa couplings of the corresponding 
generation. However all the assumptions of this type crucially depend on the high 
energy origin of the LQ models.  
Therefore,  following the common practice we keep  in our constraints both $\lambda$ and $M_{S}$
as free parameters. 

As to the mass scale $\mu$ of the trilinear operators in Eq. \rf{LQ-self},
it seems reasonable to assume that $\mu \geq \Lambda_F \sim 250$ GeV, 
where $\Lambda_F$ is electroweak scale. This could be motivated by 
the observation that these operators are associated with physics 
beyond the SM whose typical mass scale is expected to be 
larger than the electroweak scale $\Lambda_F$.
Thus, taking $\mu \sim \Lambda_F$,  we obtain a ``conservative" lower bound
\ba{M-LQ1}
M_{S} \geq \lambda^{1/2} \cdot 25\ \mbox{TeV}.
\ea 
However, the actual value of the scale $\mu$ can be much larger than $\Lambda_F$ easily reaching,
for instance, the grand unification scale $M_{GUT} \sim 10^{16}$ GeV. 
The latter case results in the constraint
\ba{M-LQ2}
M_{S} \geq \lambda^{1/2} \cdot (5\times 10^3) \ \mbox{TeV}.
\ea 
Thus the typical constraint for the case of scalar LQs ranges between those in Eq. \rf{M-LQ1}
and in Eq. \rf{M-LQ2} 
\ba{M-LQ3}
M_{S} \geq \lambda^{1/2} \cdot (25 \div 5 \times 10^3)\ \mbox{TeV}.
\ea 

The constraints for the case of vector LQs can be obtained directly from the 
Eq. \rf{M-LQ} by the substitution $\mu \rightarrow g \cdot m_p$, assuming that all the 
dimensionless coupling constants in Eq. \rf{LQ-self} are of the same order of magnitude
$g_0^i \sim \tilde{g}_0^i \sim g_1 \sim g$. This substitution is motivated by the observation
that the energy scale of the trilinear vector LQ operators in Eq. \rf{LQ-self} can be estimated 
as $g \times $[{\it energy scale of the derivative}] and that the energy scale of the 
derivative is given by the mean momentum flowing in the LQ propagators.  The latter 
is comparable with the proton mass $m_p$. Since the energy scale appears in Eq. \rf{M-LQ} 
with the small exponent 1/6, the difference between $m_p$ and the actual energy scale 
is not important. Thus for the vector LQ mass we obtain
\ba{M-LQ-V}
M_{V} \geq 
\lambda^{1/2}g^{1/6}\cdot 10\ \mbox{TeV}.
\ea

There exist in the literature constraints on LQ models from accelerator 
and non-accelerator experiments (for a summary see, for instance, 
\cite{Global-1,Global-2,Search,DBC}). The most stringent constraints for the 1st generation 
LQs follow from the measurements of Atomic Parity Violation and from the universality 
in leptonic $\pi$-decays. The best constraints from these experiments are 
\cite{Search}
\ba{APV-pi}
M_{S} \geq \lambda \cdot\ 3.5\ \mbox{TeV}, \ \ \ \ \ \ M_{V} \geq \lambda \cdot\ 6.6\ \mbox{TeV}.
\ea
The comparison of these constraints with those in Eqs. \rf{M-LQ1}-\rf{M-LQ3}
leads us to the conclusion that the proton decay constraints for the 1st generation 
scalar LQs are significantly more stringent than other existing constraints. 
For the case of vector 
LQs the comparison of the proton decay constraints with the existing ones can not 
be made in a direct way due to the presence  in Eq. \rf{M-LQ-V} of the vector 
LQ self-interaction coupling 
constant $g$. Nevertheless it is instructive to 
consider some sample values of the coupling constants $\lambda$ and $g$.  
For instance, if $\lambda \sim 1$ the proton decay constraints comparable with the constraint 
in Eq. \rf{APV-pi} occur only for large values of the vector LQ self-interaction coupling  
constant $g \geq 0.1$ which is unlikely. However for smaller values of 
the LQ-quark-lepton coupling  
$\lambda$ comparable proton decay constraints can occur at very small values of $g$. 
For instance, if $\lambda \sim 0.01$ the corresponding value of 
this coupling is $g \sim 10^{-7}$.

In conclusion, we derived new constraints on the 1st generation LQ masses
and couplings from the experimental lower bound for the proton life time. 
We have shown that for the case of the scalar LQs these constraints are more stringent 
than the corresponding constraints obtained from other experiments.

\bigskip
\noindent
{\bf ACKNOWLEDGMENTS}\\[3mm]
This work is partially supported by Fondecyt (Chile) project 8000017.

%%%%%%%%%%%%%%%%%%%%%%%%%%%%%%%%%%%%%%%%%%%%%%%%%%%%%%%%%%%%%%%%%
\newpage
\vspace*{-40mm}
\begin{table}[t]
{Table 1:
The Standard model assignments as well as lepton L and baryon B
numbers of the scalar ${ S}$ and vector ${ V}_{\mu}$  leptoquarks (LQ). 
($Y = 2(Q_{\it em} - T_3)$). 
}\\[5mm]
%\vspace*{5mm}
\begin{tabular}{|c|c|c|c|c|c|c|}
\hline
LQ & $SU(3)_c$ & $SU(2)_L$ &    $Y$    & $Q_{\it em}$ &L & B\\
\hline
\hline
%&&&&&&\\
$ S_0$         & ${\bf 3}$ & ${\bf 1}$ & -2/3 &  -1/3&1&1/3\\
\hline
%&&&&&&\\
$\tilde{S}_0$   & ${\bf 3}$ & ${\bf 1}$ & -8/3 &  -4/3&1&1/3\\
\hline
%&&&&&&\\
$S_{1/2}$    & ${\bf 3^\ast}$ & ${\bf 2}$ & -7/3  & (-2/3, -5/3)&1&-1/3\\
\hline
%&&&&&&\\
$\tilde S_{1/2}$   & ${\bf 3^\ast}$ & ${\bf 2}$ & -1/3  & (1/3, -2/3)&1&-1/3\\
\hline
%&&&&&&\\
$S_1$    & ${\bf 3}$ & ${\bf 3}$ & -2/3  & (2/3, -1/3,-4/3)&1&1/3\\
\hline
\hline
%%%%%%%%%%%%%%%%%%%%%%%%%%%%%%%%%%%%%%%%%%%%%%%%%%%%%%%%%%%
%&&&&&&\\
$ V_0$         & ${\bf 3^\ast}$ & ${\bf 1}$ & -4/3 &  -2/3&1&1/3\\
\hline
%&&&&&&\\
$\tilde{V}_0$   & ${\bf 3^\ast}$ & ${\bf 1}$ & -10/3 &  -5/3&1&1/3\\
\hline
%&&&&&&\\
$V_{1/2}$    & ${\bf 3}$ & ${\bf 2}$ & -5/3  & (-1/3, -4/3)&1&-1/3\\
\hline
%&&&&&&\\
$\tilde V_{1/2}$   & ${\bf 3}$ & ${\bf 2}$ & 1/3  & (2/3, -1/3)&1&-1/3\\
\hline
%&&&&&&\\
$V_1$    & ${\bf 3^\ast}$ & ${\bf 3}$ & -4/3  & (1/3, -2/3,-5/3)&1&1/3\\
\hline
\end{tabular}
\end{table}
%%%%%%%%%%%%%%%%%%%%%%%%%%%%%%%%%%%%%%%%%%%%%%%%%%%%%%%%%%%%%%%%%%%%%
\end{document}